\documentclass[prl,showpacs,twocolumn]{revtex4}

\usepackage{graphicx}
\usepackage{amsmath}
\usepackage{slashed}
\usepackage{feynmp}

\begin{document}
\title{On explaining the observed pattern of quark and lepton masses}
\author{Ji\v r\'{\i} Ho\v sek}
\email{hosek@ujf.cas.cz} \affiliation{Department of Theoretical
Physics, Nuclear Physics Institute, Czech Academy of Sciences, 25068
\v Re\v z (Prague), Czech Republic}

\begin{abstract}
\noindent Higgs sector of the Standard model (SM) is replaced by the gauge $SU(3)_f$ quantum flavor dynamics (QFD) with
one parameter, the scale $\Lambda$. Anomaly freedom of QFD demands extension of the fermion sector of SM by three sterile right-handed neutrino fields.
Poles of fermion propagators with chirality-changing self-energies $\Sigma(p^2)$ spontaneously generated by QFD at strong coupling define:
(1) Three sterile-neutrino Majorana masses $M_{fR}$ of order $\Lambda$. (2) Three Dirac masses $m_f$, degenerate for $e_f, \nu_f, u_f, d_f$ in family $f$,
exponentially small with respect to $\Lambda$. Goldstone theorem implies: All eight flavor gluons acquire
masses of order $M_{fR}$. $W$ and $Z$ bosons acquire masses of order $\sum m_f$, the effective Fermi scale.
Composite 'would-be' Nambu-Goldstone bosons have their 'genuine' partners, the composite Higgs particles:
The SM-like Higgs $h$ and two new Higgses $h_3$ and $h_8$, all with masses at Fermi scale;
three Higgses $\chi_i$ with masses at scale $\Lambda$. Large pole-mass splitting of charged leptons and quarks in $f$ is
arguably due to full QED $\Sigma(p^2)$-dependent fermion-photon vertices enforced by Ward-Takahashi identities.
The argument relies on illustrative computation of pole-mass splitting found non-analytic in fermion electric charges.
Neutrinos are the Majorana particles with seesaw mass spectrum computed solely by QFD.
Available data fix $\Lambda$ to, say, $\Lambda \sim 10^{14} \rm GeV$.
\end{abstract}

\maketitle

\section{I. Introduction and summary}
Lack of explanation of the observed wide and wild charged lepton and
quark mass spectrum is a nightmare of theoretical elementary
particle physics \cite{weinberg1}. The extreme smallness of neutrino masses manifested
in neutrino oscillations is its almost unbearable stage. Elegant description of fermion massiveness,
however, does exist. The charged lepton and quark masses are
described in the SM model by the essentially classical Higgs
mechanism \cite{higgs},\cite{englert-brout}, and the extreme lightness of neutrinos is
described in its minimal extension by the
entirely classical seesaw \cite{seesaw}.

The calculable mass spectrum of {\it fermion quantum fields} viewed as
coupled quantum oscillators is conceivable by replacing the Higgs mechanism by the dynamical symmetry
breakdown. First heuristic model of this sort was suggested by Yoichiro Nambu \cite {njl}. We refer with admiration to the papers
of Heinz Pagels \cite{pagels-stokar}, \cite{carter-pagels}, \cite{pagels}: These papers
analyze some consequences of the dynamical Higgs mechanism which he calls quantum flavor dynamics (QFD) \cite{pagels-stokar}.
The main points of his QFD, dealing with SM fermions and some gauge fields, without specifying the Lagrangian, are common with our approach:
First, the dynamically generated fermion masses are finite and calculable \cite{pagels-stokar}. Second, their dynamical, spontaneous
generation implies the masses of $W$ and $Z$ bosons \cite{carter-pagels}. Third, there is a composite Higgs particle \cite{carter-pagels}.
We find the name QFD very appropriate and take the liberty of using it for our gauge flavor dynamics.
The first authors advocating spontaneous electroweak gauge symmetry breakdown
without elementary scalar fields were apparently Francois Englert and Robert Brout \cite{eb}.
Other approaches, which have to be quoted in the present context are the
'technicolor'(TC) \cite{weinberg}, \cite{susskind} and the 'extended technicolor' (ETC) \cite{etc}.
They differ from our approach by the need of new symmetries and new fermions.

We suggest \cite{hosek} to employ the gauge flavor (family, generation, horizontal) dynamics
defined by the gauge $SU(3)_f$ invariant Lagrangian
\begin{eqnarray*}
{\cal L}_f=-\frac{1}{4}F_{a\mu\nu}F_a^{\mu\nu}+ \ \bar q_L i\slashed
D q_L + \bar u_R i\slashed D u_R + \bar d_R i\slashed D d_R\\ + \bar
l_L i\slashed D l_L + \bar e_R i\slashed D e_R + \bar \nu_{R}
i\slashed D \nu_{R}
\end{eqnarray*}
The field tensor $F$ describes the kinetic term of eight flavor gluons $C$
and their self-interactions, the covariant derivatives $D$ describe
their interactions with chiral, both right- and left-handed, fermions. The $q_L$ and $l_L$ are the
quark and lepton electroweak doublets, respectively, $u_R, d_R, e_R,
\nu_R$ are the electroweak singlets. Their weak hypercharges are uniquely fixed by the corresponding
electric charges.

All chiral fermion fields transform as triplets of flavor, or horizontal,
or family or generation $SU(3)_f$ symmetry.
The theory defined by ${\cal L}_f$ is by construction anomaly free.
The gauge $SU(3)_f$ interaction, characterized by one dimensionless coupling
constant $h$, is asymptotically free at high momenta and hence
strongly interacting in the infrared. This means that by dimensional
transmutation the dimensionless $h$ can turn into the theoretically
arbitrary mass scale $\Lambda$.

It is the firm experimental fact that the $SU(3)_f$ symmetry
manifest in the Lagrangian ${\cal L}_f$ is observed as badly broken. The crucial
theoretical question therefore is  whether the gauge $SU(3)_f$
dynamics is capable of the observed spontaneous self-breaking.

The Lagrangian ${\cal L}_f$ is considered together with the standard electroweak gauge $SU(2)_L
\times U(1)_Y$ forces known to remain weakly coupled all the way up to
the Planck scale. The QCD is present as a spectator.

We admit that the idea of spontaneous self-breaking of the flavor $SU(3)_f$ symmetry is nasty and suspicious \cite{vafa-witten}:
The Lagrangian ${\cal L}_f$ is formally identical with the QCD Lagrangian.
Since in QCD we trust we should provide a truly good reason why in
the infrared the suggested QFD spontaneously self-breaks whereas the QCD confines.

We are convinced there is such a good reason: While the QCD deals with the
electrically charged quarks, the QFD deals also with the
electrically neutral sterile right-handed neutrinos {\it which can be the massive Majorana particles}.
It is utmost important that their hard mass term
\begin{equation}
{\cal L}_{Majorana}=-\tfrac{1}{2}(\bar \nu_{R}M_{R}(\nu_{R})^{{\cal
C}}+ h.c.)\label{majorana}
\end{equation}
{\it unlike the Dirac mass term}, is strictly prohibited by the $SU(3)_f$ gauge symmetry: It transforms as $3^{*} \times 3^{*} = 3 + 6^{*}$
which does not contain unity. It can, however, be generated dynamically \cite{njl} provided this option is energetically
favorable. It is easy to see that the relevant part of the Lagrangian
\begin{eqnarray*}
{\cal L}_{\rm int}=\tfrac{1}{2}h\{\bar
\nu_{R}\gamma_{\mu}\tfrac{1}{2}\lambda_a\nu_{R}+ \overline
{(\nu_{R})^{{\cal
C}}}\gamma_{\mu}[-\tfrac{1}{2}\lambda_a^T](\nu_{R})^{{\cal
C}}\}C^{\mu}_a
\end{eqnarray*}
in contrast with the vector-like QCD Lagrangian, is effectively chiral:
The charge conjugate neutrino field $(\nu_R)^{{\cal C}}= C (\bar \nu_R)^T$
is of course a {\it left-handed field}, $(\nu_R)^{{\cal C}}=(\nu^{\cal
C})_L$. Unlike the other left-handed fields in the Lagrangian ${\cal L}_f$ it transforms, however, as the {\it antitriplet} of $SU(3)_f$:
$T_a(L)=-\tfrac{1}{2}\lambda_a^{T}$. Harsh support or invalidation of our suggestion can thus apparently be given only
by the non-perturbative lattice computations.\\

(I) Fermion mass is a bridge between the right- and the left-handed fermion field. In the
Lagrangian under consideration {\it all hard fermion mass terms} are strictly prohibited by its  gauge $SU(3)_f \times SU(2)_L \times U(1)_Y$ invariance \cite{argument}.
We demonstrate below that the QFD dynamics itself, strongly coupled in the infrared,  generates spontaneously:
(i) Three Majorana masses $M_{fR}$ of sterile right-handed neutrinos of order $\Lambda$. This, once for ever fixes $\Lambda$. The conservative value is, say $\Lambda \sim 10^{14}\rm GeV$.
We believe that, if necessary, it could be fixed even to the Planck scale \cite{donoghue}.
(ii) In identical approximation the QFD dynamics generates three Dirac masses $m_f$ exponentially small with respect to $\Lambda$. {\it Because there is nothing
in QFD which would distinguish within family $f$ between the neutrino, the charged lepton, the $Q=2/3$ quark, and the $Q=-1/3$ quark,
the masses $m_f$ must come out degenerate}. They are obtained in a crude separable approximation to the kernel
of the Schwinger-Dyson (SD) equation for chirality-changing fermion self-energies $\Sigma_f(p^2)$. Big advantage is the resulting explicit
form of the Euclidean $\Sigma_f(p^2) = m_f^2/p$. The fermion masses $M_{fR}$ and $m_f$
are defined as poles of the full fermion propagators $S(p)=(\slashed p - \Sigma_f(p^2))^{-1}$.

{\it We assume that the fermion mass pattern $M_{fR} \gg m_f$ obtained in a crude approximation is the generic property of QFD
at strong coupling}. It then follows that it fixes the spontaneous symmetry-breaking pattern
of the underlying gauge $SU(3)_f \times SU(2)_L \times U(1)_Y$ symmetry uniquely, and allows to draw several strong conclusions:
For transparency we rewrite $m_f$ as $m_f \equiv m_{(0)}\lambda_0 + m_{(3)} \tfrac{1}{2}\lambda_3 + m_{(8)} \tfrac{1}{2}\lambda_8$ where
\begin{eqnarray*}
m_{(0)} &=& \tfrac{1}{\surd 6}(m_1 + m_2 + m_3)\\
m_{(3)} &=& m_1 - m_2\\
m_{(8)} &=& \tfrac{1}{\surd 3}(m_1 + m_2 - 2m_3)
\end{eqnarray*}\\

(II) (1) The Majorana masses $M_{fR}$ break down the gauge symmetry $SU(3)_f$ spontaneously and completely \cite{yanagida}, \cite{bhs}, \cite{hosek}. Consequently, eight
'would-be' NG bosons composed of sterile neutrinos give rise to masses $m_{iC}$ of all flavor gluons $C$ of order $\Lambda$ \cite{migdal-polyakov}, \cite{jackiw-johnson}, \cite{cornwall-norton}.\\

(2) Eight 'would-be' NG bosons and one genuine pseudo NG boson resulting from spontaneous breakdown of global
anomalous $U(1)$ symmetry of the sterile neutrino sector belong to the complex composite sextet \cite{hosek}
\begin{equation*}
\Phi_{fg} \sim (\bar \nu_{fR}(\nu_{gR})^{{\cal C}})
\end{equation*}
Consequently, as a remnant of symmetry ( $8 + 1 + 3 = 12$), there should exist three genuine Higgs-like composite bosons $\chi_i$ with masses of order  $\Lambda$ \cite{hosek}.\\

(III) (1) The Dirac mass $m_{(0)}$ breaks down the electroweak gauge $SU(2)_L \times U(1)_Y$ symmetry spontaneously down to $U(1)_{em}$,
and leaves the flavor $SU(3)_f$ gauge symmetry intact. Consequently, three multi-component 'would-be' NG bosons composed of all electroweakly interacting
leptons and quarks give rise to masses $m_W$ and $m_Z$ of $W$ and $Z$ bosons, respectively, in terms of $\sum m_f$. As the fermion masses in families come out
degenerate the canonical Weinberg relation $m_W/m_Z = \rm cos \theta_W$ is exact. These masses, which define the {\it induced electroweak scale} are related to
the fermion masses by sum rules.\\

(2) Three 'would-be' NG bosons belong to the complex multi-component composite doublet (index $a$ in the following formula)
\begin{equation*}
\phi^a \sim (\bar \psi_{fR} \psi^{af}_{L})
\end{equation*}
($\psi_{L,R}$ are the electroweakly interacting chiral SM fermion fields). Consequently, as a remnant of symmetry (as in the Standard model)
($3 + 1 = 4$), there should exist one genuine multi-component composite SM-like Higgs boson $h$ with mass at the electroweak scale.\\

(IV) (1) The Dirac masses $m_{(3)}$ and $m_{(8)}$ break down spontaneously the flavor $SU(3)_f$ gauge symmetry down to $U(1) \times U(1)$ \cite{higgs}.
Consequently, six multi-component 'would-be' NG bosons composed of the electroweakly interacting leptons and quarks contribute a tiny amount
to huge masses of six flavor gluons \cite{higgs}. These contributions can be safely neglected.\\

(2) Six multi-component 'would-be' NG bosons belong to the real octet (index $i$ in the following formula)
\begin{equation*}
\phi_i \sim  [(\bar \psi_{fR} (\lambda_i)^f_g \psi^g_L) + h.c.]
\end{equation*}
($\psi_{L,R}$ are the electroweakly interacting chiral SM fermion fields). Consequently, as a remnant of symmetry ($6 + 2 = 8$)
there should exist two additional multi-component composite Higgs-like bosons $h_3$ and $h_8$ with masses at the electroweak scale. It is rather remarkable that
namely such a possibility was explicitly mentioned as an example of the non-Abelian Higgs mechanism by Peter Higgs in his seminal paper \cite{higgs}.\\

(V) The Dirac masses $m_{(3)}$ and $m_{(8)}$ break down spontaneously also the electroweak gauge $SU(2)_L \times U(1)_Y$ symmetry spontaneously down to $U(1)_{em}$.
Consequently, there are additional components of the composite 'would-be' NG bosons and of the genuine Higgs particle $h$.
Detailed analysis is left for future work.\\

(VI) The observed fermion mass spectrum in families is not degenerate but widely split. Inspired by the prescient Heinz Pagels \cite{pagels-stokar}
we conjecture that the observed {\it large} fermion mass splitting of charged leptons and quarks in families is due
to QED. This on the first sight absurd explanation is conceivable due to the following:
The electromagnetic contributions to the fermion propagators with $\Sigma_f(p^2)$ are not the standard QED corrections.
First of all, due to $\Sigma_f(p^2)=m_f^2/p$ they are UV finite. Moreover, position of particle poles in fermion propagators with complicated functions $\Sigma(p^2)$
is not a priori the perturbative notion. Finally, and most important,
there are entirely new chirality-changing fermion-photon vertices {\it enforced} in the electromagnetic Ward-Takahashi (WT) identities
by the momentum-dependent self-energies $\Sigma(p^2)$ \cite{pagels-stokar}, \cite{ball-chiu}, \cite{delbourgo}, \cite{benes-hosek}. It is gratifying that
the illustrative computation of the UV finite QED contribution \cite{pagels-stokar} with $\Sigma(p^2) = m^2/p$ and with the new fermion-photon vertex
\begin{equation*}
\Gamma^{\mu}(p',p)=\gamma^{\mu}-(p'+p)^{\mu}[\Sigma(p'^{2})-\Sigma(p^2)]/(p'^{2}-p^2)
\end{equation*}
can be done {\it exactly}, and that it supports the conjecture.\\

The neutrino electric charge is zero and, consequently, the neutrino mass spectrum is predicted solely by QFD in terms of $m_f$ and $M_{fR}$ by seesaw \cite{hosek-seesaw}.
The available data restrict the scale $\Lambda$ to  $\Lambda \sim 10^{14} \rm GeV$.\\

(VII) There are interesting phenomena associated with spontaneous breakdown of global chiral Abelian symmetries of the model \cite{hosek}.
The anomalous ones result in observable axion-like particles, the anomaly-free one can be gauged, resulting in
new massive  $Z'$ gauge boson. These topics are not discussed in the present paper.\\

In 1979 Tsutomu Yanagida published the electroweak quantum field theory with naturally incorporated seesaw \cite{yanagida}: First, introduction of one triplet
of sterile right-handed neutrinos enforced by the quantum-field theoretic requirement of anomaly freedom of the gauged flavor $SU(3)_f$ symmetry
of three families of the SM chiral fermion fields is natural. Moreover, gauging the flavor symmetry is natural by itself. Not surprisingly it was attempted
by notable authors \cite{gauging-flavor}. Second, since the gauge $SU(3)_f \times SU(2)_L \times U(1)_Y$ gauge symmetry does not tolerate any hard fermion mass terms
(unlike the phenomenological seesaw) Yanagida was forced to introduce appropriate weakly coupled elementary Higgs fields for their generation.
In conclusion of \cite{yanagida} Yanagida notes 'that the model is a possible candidate for the spontaneous mass generation by dynamical symmetry breaking'.
Our strong-coupling results presented here and elsewhere \cite{hosek}
can thus be compared with theoretically safe but phenomenological weak-coupling results of Yanagida.\\

In Sect.II we describe the explicit approximate solution of the QFD Schwinger-Dyson (SD) equation for the chirality-changing
fermion self energies $\Sigma_f(p^2)$ which yield
$M_{fR}$ and $m_f$. These masses generated by the strongly coupled QFD define uniquely the pattern of spontaneously broken symmetries.
As a consequence the same strongly coupled QFD is responsible for the formation of both composite 'would-be' NG bosons
and of the composite genuine Higgs particles. In Sect.III we illustrate the possibility of large charge-lepton and quark mass splitting due to the QED
contributions to $\Sigma_f(p^2)$ due to entirely new fermion-photon vertices.
This mass splitting is not accompanied by any additional spontaneous symmetry breaking. It is then to be expected that
the gauge electroweak corrections to the masses of the gauge bosons and to basic properties of the Higgs particles will
only be perturbative, as in the SM \cite{carter-pagels}. In the concluding Sect.IV we briefly put the obtained results into a wider context
and point out a natural QFD candidate for dark matter.\\

\section{II. QFD origin of fermion masses}
Our aim is to find the non-perturbative solutions of the SD equation for the chiral-symmetry-breaking $\Sigma(p^2)$ in
the full fermion propagators $S^{-1}(p)=\slashed p - \Sigma(p)$
for all fermions of the $SU(3)_f \times SU(2)_L \times U(1)_Y$ gauge-invariant Lagrangian.
It is important to realize that the SD equation of QFD presented below (\ref{Sigma}) is universal: (i) Majorana mass is the R-L bridge between
$\nu_R$ which transforms as a flavor triplet, and between the left-handed  $(\nu_R)^{{\cal C}}$
which transforms as flavor anti-triplet. The corresponding $\Sigma(p^2)$ which gives rise to Majorana masses is the complex $3 \times 3$ matrix,
the symmetric sextet by Pauli principle, and $T_a(R)=\tfrac{1}{2}\lambda_a$,   $T_a(L)=-\tfrac{1}{2}\lambda_a^T$.
(ii) Dirac mass is the R-L bridge between the right- and left-handed fermion fields
both transforming as flavor triplets: The corresponding $\Sigma(p^2)$ which gives rise to Dirac masses is a general complex $\bar 3 \times 3$ matrix,
and $T_a(R)=T_a(L)=\tfrac{1}{2}\lambda_a$. Hence there is nothing in QFD which would distinguish between
the Dirac masses of the neutrino, the charged lepton, the charge $Q=2/3$ quark, and the charge $Q=-1/3$ quark in given family. Difference between
the Majorana and Dirac mass matrices turns out, however, substantial.

Moreover, we are searching for the UV-finite symmetry-breaking solutions $\Sigma$, because the fermion mass counter terms are prohibited \cite{pagels-stokar}. The
Schwinger-Dyson equation is a homogeneous non-linear integral equation which has the form \cite{pagels}, \cite{hosek}
\begin{widetext}
\begin{eqnarray}
\Sigma(p)=3\int \frac{d^4k}{(2\pi)^4}\frac{\bar
h^2_{ab}((p-k)^2)}{(p-k)^2}T_a(R)\Sigma(k)[k^2+\Sigma^{+}(k)\Sigma(k)]^{-1}T_b(L)
\label{Sigma}
\end{eqnarray}
\end{widetext}

According to the Nambu's self-consistent reasoning \cite{njl} we first {\it assume} that the gauge flavor $SU(3)_f$ is completely
self-broken, and subsequently {\it find} the corresponding symmetry-breaking solutions.

The sliding coupling $\bar h_{ab}^2(q^2)$ in (\ref{Sigma}) defined
in terms of the flavor gluon polarization tensor contains important
information about the assumed low-momentum properties of the model.
In particular, it corresponds to the phase in which all flavor
gluons are massive. Despite this, it remains unknown. The reason is
that the spectrum of the expected composites carrying flavor, which
by definition below $\Lambda$ contribute to $\bar h_{ab}^2(q^2)$, is
entirely unknown. Finding the fermion mass spectrum is therefore a
formidable task.

In order to proceed we approximate the problem as follows:

(1) In the perturbative weak coupling high-momentum region from
$\Lambda$ to $\infty$ which in technical sense guarantees the UV
finiteness of $\Sigma(p^2)$ \cite{pagels} we set the known \cite{af}
perturbative i.e. small, $\bar h_{ab}^2(q^2)$
\begin{equation*}
\frac{\bar h_{ab}^2(q^2)}{4\pi}=\frac{\delta_{ab}}{(11-\tfrac{n_f}{3})\rm
ln(q^2/\Lambda^2)}
\end{equation*}
equal to zero. Here $n_f=16$ is the number of chiral fermion
triplets in the model. {\it The resulting model is thus not asymptotically, but
strictly free above the scale $\Lambda$.}

(2) Without loss of generality we fix in the resulting SD equation the external
euclidean momentum as $p=(p,\vec 0)$, integrate over angles and get
\begin{equation}
\Sigma(p)=\int_0^{\Lambda}k^3dk
K_{ab}(p,k)T_a(R)\Sigma(k)[k^2+\Sigma^{+}\Sigma]^{-1}T_b(L)
\label{Sigmasep}
\end{equation}
Here the {\it unknown} kernel
\begin{equation}
K_{ab}(p,k)\equiv \frac{3}{4\pi^3}\int_0^{\pi}\frac{\bar
h_{ab}^2(p^2+k^2-2pk \rm \cos \theta)}{p^2+k^2-2pk \rm \cos
\theta}\rm \sin^2 \theta d \rm \theta
\end{equation}
is separately symmetric in momenta and in the flavor octet indices.

(3) Our key approximation is the {\bf separable approximation} for
the kernel $K_{ab}(p,k)$. In the following we analyze explicitly the
Ansatz
\begin{equation}
K_{ab}(p,k)=\frac{3}{4\pi^2} \frac{g_{ab}}{pk}\label{sep}
\end{equation}

The Ansatz mimics our ignorance of knowing the low-momentum
$\bar h_{ab}^2(q^2)$ {\it and} the low-momentum form of the flavor
gluon propagators (to be found subsequently). Ultimately we should
deal with a system of Schwinger-Dyson equations for several Green
functions, an entirely hopeless task.

Here $g_{ab}$ is an appropriate real symmetric numerical matrix of the effective
low-momentum dimensionless coupling constants which in the present approximation
reflect the complete breakdown of $SU(3)_f$.

Our motivation for the separable approximation (approved eventually
a posteriori) is the following:

1. The nonlinearity of the integral equation is preserved. We expect
that the non-analyticity of $\Sigma$ upon the couplings $g_{ab}$ is
crucial for generating the huge fermion mass ratios.

2. In separable approximation the homogeneous nonlinear integral
equation (\ref{Sigmasep}) is immediately formally solved:
\begin{equation}
\Sigma(p)=\frac{\Lambda^2}{p}T_a(R)\Gamma_{ab}T_b(L)\equiv\frac{\Lambda^2}{p}\sigma
\label{sol}
\end{equation}
The difficult part is that the numerical matrix $\Gamma$ has to
fulfil the nonlinear algebraic self-consistency condition (gap
equation)
\begin{eqnarray}
\Gamma_{ab}&=&g_{ab}\frac{3}{16\pi^2}\int_{0}^{1}dx (T(R)\Gamma
T(L)) \nonumber\\
&& [x + (T(R)\Gamma T(L))^{+}(T(R)\Gamma T(L))]^{-1}\label{Gamma}
\end{eqnarray}

3. The assumed momentum dependence of $\Sigma(p)\sim 1/p$ is not
without support. Because the masses of flavor gluons come out huge
it is justified to think heuristically of the dynamically generated
fermion masses in terms of the four-fermion interaction of Nambu and
Jona-Lasinio \cite{njl}. In a series of papers \cite{mannheim}
Philip Mannheim argues that the theoretically  consistent treatment
of the fermion mass generation by the four-fermion dynamics should
result namely in $\Sigma(p) \sim 1/p$. It is gratifying that the behavior $\Sigma(p)\sim 1/p$
does not imply any infrared divergences. It merely modifies the denominators
of the fermion propagators in the Euclidean loop integrals into the form $p^4 + m^4$.

4. There is the deep analogy between the spontaneously broken phase symmetry in non-relativistic superconductors resulting
in non-zero gap $\Delta$ in the dispersion law of quasi-electrons, and the spontaneously broken chiral symmetry
in the Lorentz-invariant Standard model resulting in non-zero fermion masses $M_{fR}$ and $m_f$.
Why the simple separable BCS approximation taking into account only the opposite
momenta around Fermi surface is phenomenologically so successful  was a mystery for years. Its theoretical relevance
was clarified much later by Polchinski \cite{polchinski}. We believe that the separable Ansatz resulting in the
phenomenologically appealing fermion mass spectrum will also be theoretically justified.

{\bf For neutrinos} $\Sigma$ describes both the masses of Majorana
neutrinos, and their mixing (including the new CP-violating phases):
The general complex symmetric $3 \times 3$ matrix $\sigma$ can be
put into a positive-definite real diagonal matrix $\gamma$ by a
constant unitary transformation
\begin{equation}
\sigma=U^{+}\gamma U^{*}
\end{equation}
The gap equation becomes
\begin{equation}
\gamma=UT_a(R)U^{+}g_{ab}I(\gamma)U^{*}T_b(L)U^{T} \label{gammaM}
\end{equation}
where
\begin{eqnarray}
I(\gamma)&=&\frac{3}{16\pi^2}\gamma
\int_{0}^{1}\frac{dx}{x+\gamma^2}= \frac{3}{16\pi^2}\gamma \rm
ln\frac{1+\gamma^2}{\gamma^2}\nonumber\\
%&\approx& -\frac{3}{16\pi^2}\gamma \rm ln \gamma^2
\end{eqnarray}

The diagonal entries of the equation (\ref{gammaM}) determine the
sterile neutrino masses, the nondiagonal entries provide relations
for the mixing angles and the new CP-violating phases. These phases
are most welcome as a source of an extra CP violation needed for
understanding of the baryon asymmetry of the Universe
\cite{leptogenesis}.

{\bf For Dirac fermions}, as in the Standard model, the generally
complex $3 \times 3$ matrix $\sigma$ can be put into a
positive-definite real diagonal matrix $\gamma$ by a constant
bi-unitary transformation:
\begin{equation}
\sigma=U^{+}\gamma V
\end{equation}
The gap equation becomes
\begin{equation}
\gamma=UT_a(R)U^{+}g_{ab}I(\gamma)VT_b(L)V^{+} \label{gammaD}
\end{equation}
The diagonal entries of the equation (\ref{gammaD}) determine the
fermion masses, the nondiagonal entries provide relations for the
CKM mixing angles and the SM CP-violating phase.

In the following we neglect the fermion mixing i.e., set the mixing matrices in (13) and (16)
equal to one. The gap equations take the form for Dirac and Majorana masses, respectively
\begin{equation}
\gamma=\tfrac{1}{4}\lambda_a g_{ab} I(\gamma)\lambda_b\label{A1}
\end{equation}
\begin{equation}
\gamma=-\tfrac{1}{4}\lambda_a g_{ab}
I(\gamma)\lambda_b^{T}\label{A2}
\end{equation}
Here $\lambda_a$ are the Gell-Mann matrices.
Because $\Sigma(p^2)\equiv
\tfrac{\Lambda^2}{p}\gamma$, the fermion mass, defined as a pole of
the full fermion propagator is
\begin{equation*}
m \equiv \Sigma(p^2=m^2)=\Lambda \gamma^{1/2}
\end{equation*}

With $g_{11}, g_{22}, g_{33}, g_{38}, g_{44}, g_{55}, g_{66},
g_{77}, g_{88}$ different from zero the right hand sides of
equations (\ref{A1}) and (\ref{A2}) are the diagonal matrices. The
equations themselves can be rewritten as
\begin{equation}\label{A3}
\gamma_i^{D/M}=\sum_{k=1}^3 \alpha_{ik}^{D/M}\gamma_k^{D/M}
\ln\frac{1+(\gamma_k^{D/M})^2}{(\gamma_k^{D/M})^2}
\end{equation}
where
\begin{widetext}
\begin{equation}\label{alpha}
\alpha^{D/M}=\frac{3}{64\pi^2}\left(\begin{array}{ccc}
\pm\left(g_{33}+\frac{2}{\sqrt{3}}g_{38}+\frac{1}{3}g_{88}\right) & g_{22}\pm g_{11} & g_{55}\pm g_{44}\\
g_{22}\pm g_{11} & \pm\left(g_{33}-\frac{2}{\sqrt{3}}g_{38}+\frac{1}{3}g_{88}\right) & g_{77}\pm g_{66}\\
g_{55}\pm g_{44} & g_{77}\pm g_{66} & \pm \frac{4}{3}g_{88}
\end{array}
\right)
\end{equation}
\end{widetext}
and the upper/lower signs correspond to the Dirac fermion masses
and the Majorana neutrino masses, respectively.

\subsection{1. Dirac masses $m_f$}
The case of Dirac masses is simpler and will be analyzed first. The equation (\ref{alpha}) suggests further
simplification, and we consider only
\begin{equation*}
g_{33},g_{38},g_{88};g_{11}=-g_{22},g_{44}=-g_{55},g_{66}=-g_{77}
\end{equation*}
different from zero.

The matrix gap equation for the Dirac masses $m_i$ becomes
diagonal and decoupled, and it is easily solved. Provided the
combinations
\begin{eqnarray*}
\alpha_{11}&=&\tfrac{3}{64\pi^2}(g_{33}+\tfrac{2}{\sqrt
3}g_{38}+\tfrac{1}{3}g_{88})\\
\alpha_{22}&=&\tfrac{3}{64\pi^2}(g_{33}-\tfrac{2}{\sqrt
3}g_{38}+\tfrac{1}{3}g_{88})\\
\alpha_{33}&=&\tfrac{3}{64\pi^2}\tfrac{4}{3}g_{88}
\end{eqnarray*}
are all positive and all $\alpha_{ii} \ll 1$, the resulting explicit Dirac
mass formulas are
\begin{equation}
m_i=\Lambda \phantom{b} \rm exp \phantom{b}
(-1/4\alpha_{ii})\label{mD}
\end{equation}
well suited for parameterizing large mass splitting between families.
\vspace{3mm}

\subsection{2. Majorana masses $M_{fR}$}
First we note that for $g_{11}=g_{44}=g_{66}=0$, the gap
equations for the Majorana masses would have no solution because of the
minus sign in front of the $\alpha_{ii}$. Consequently, $(g_{11},
g_{44}, g_{66})\neq 0$. Second, in the case of sterile Majorana
neutrinos we are not aware of the necessity of the hierarchical mass
spectrum. The phenomenological constraint which we want to respect is
\begin{equation*}
M_{fR} \sim \Lambda
\end{equation*}
With the constants $\alpha_{ii}$ fixed by the numerical
values of the Dirac masses the equations (\ref{A3}) for $\gamma_i^M$
can be viewed as a system of three inhomogeneous linear equations
for the unknown $(g_{11},g_{44},g_{66})$:
\begin{widetext}
$$
-\frac{1}{2}\left(\begin{array}{ccc}
I(\gamma_2^M)&I(\gamma_3^M)&0\\
I(\gamma_1^M)&0& I(\gamma_3^M)\\
0&I(\gamma_1^M)&I(\gamma_2^M)
\end{array}\right)
\left(\begin{array}{c} g_{11} \\g_{44} \\g_{66}
\end{array}\right)=
\left(\begin{array}{c}
\gamma_1^M+\tfrac{16\pi^2}{3}\alpha^{D}_{11} I(\gamma_1^M)\\
\gamma_2^M+\tfrac{16\pi^2}{3}\alpha^{D}_{22} I(\gamma_2^M)\\
\gamma_3^M+\tfrac{16\pi^2}{3}\alpha^{D}_{33} I(\gamma_3^M)
\end{array}\right).
$$
\end{widetext}
This set of equations has a solution for any set of $\gamma_i^M >0$.

It is important that the precise size and hierarchy of $\gamma_i^D$ does not play any important role for the numerical values of
$\gamma_i^M$. Numerical illustration enables to conclude that the universal SD equation has as its solutions both the huge Majorana masses $M_{fR}$ of sterile neutrinos
and the hierarchical, exponentially small Dirac masses $m_f$ common to all fermion species in family $f$.\\

At this point it is appropriate to raise the question of how many free parameters are
{\it ultimately} necessary for computing the fermion masses in QFD.
{\it The problem of the fermion mass spectrum is the problem of the internal (bound-state) structure of the underlying
composite fermion-antifermion condensates}:

(i) In the case of Majorana masses of sterile neutrinos it is the condensate
\begin{equation*}
\Phi_{fg} \sim (\bar \nu_{fR}(\nu_{gR})^{{\cal C}}),
\end{equation*}
the composite flavor sextet. The phenomenological description of its condensation in terms of the Higgs sextet $\Phi_{fg}$ \cite{bhs}
deals with three algebraically independent condensates $\rm tr\Phi^{+}\Phi$, $\rm tr(\Phi^{+}\Phi)^2$, and $\rm det\Phi^{+}\Phi$.

(ii) In the case of Dirac masses of electroweakly interacting fermions $l_L, q_L, \nu_R, e_R, u_R, d_R$ there are two types of the
composite multi-component condensates. One flavorless electroweak doublet $\phi$
\begin{equation*}
\phi \sim (\bar e_R l_L), (\bar d_R q_L); \tilde {\phi}\equiv (i\tau_2)\phi^{*} \sim (\bar \nu_R l_L), (\bar u_R q_L)
\end{equation*}
and one flavor octet $\phi_i$
\begin{equation*}
\phi_{i} \sim (\bar e \lambda_i e), (\bar d \lambda_i d),(\bar \nu  \lambda_i \nu), (\bar u  \lambda_i u)
\end{equation*}
The phenomenological description of their condensation in terms of the elementary Higgs fields $\phi$ and $\phi_{i}$
deals also with three algebraically independent condensates: $\phi^{+}\phi$, $\phi^i \phi_i$ and $\phi^i \phi^j \phi^k d_{ijk}$. Here $d_{ijk}$
is the cubic Casimir operator of $SU(3)$.

The strong non-Abelian $SU(3)_f$ dynamics is characterized by one theoretically arbitrary parameter, the scale $\Lambda$. Consequently,
provided our basic assumption of the complete self-breaking is warranted both the Majorana fermion masses $M_{fR}$ and the
Dirac masses $m_f$ should ultimately be the {\it calculable} multiples of $\Lambda$ \cite{pagels-stokar}. The belief here is entirely analogous
to the belief in understanding the hadron mass spectrum of the confining QCD in the chiral limit: With one
theoretically arbitrary scale $\Lambda_{QCD}$ there are the massless NG pions,
whereas the masses of all other hadrons are ultimately the calculable (so far only by a computer) multiples of $\Lambda_{QCD}$.
The case of QFD is even more complex because besides the masses of its elementary excitations (leptons, quarks and flavor gluons) there are also
the masses of its expected unconfined but strongly coupled collective excitations.

The results presented above are modest. We appreciate that with six parameters of similar order of magnitude
the SD equation generates both huge Majorana masses of sterile neutrinos, and the exponentially light Dirac masses
of the electroweakly interacting fermions.\\

\section{III. QED origin of fermion mass splitting}
{\it The electroweak gauge interactions do distinguish between different Dirac fermion species within families without any new parameters}:
First, by different electric charges. Second, by different couplings with massive $Z$ - bosons.
Because the electroweak interactions stay weakly coupled all the way up to the Planck scale
the mass splitting due to them is not accompanied by any extra spontaneous symmetry breaking. Both the 'would-be' NG bosons and the Higgs particles
are the consequences of spontaneous generation of $M_{fR}$ and $m_f$ by the strongly coupled QFD. We will argue below that the QED contributions
to fermion propagators with $\Sigma_f(p^2)$ can nevertheless be large as observed essentially due to entirely new
chirality-changing $\Sigma_f(p^2)$-induced fermion photon vertices enforced by the QED WT identities.\\

Self-energies $\Sigma_f(p^2)$ play the pivotal role in our program. First, they give rise to $M_{fR}$ and $m_f$.
Second, in the {\it axial-vector} WT identities they visualize spontaneous chiral symmetry breaking
as residues at the massless poles of the composite 'would-be' NG bosons. Masses of the gauge bosons are then computed in terms of them by the Pagels-Stokar formula.\\

It is intriguing that the self-energies $\Sigma_f(p^2)$ of electroweakly interacting fermions play the important role also
in the {\it vectorial} WT identities not associated with any obvious breakdown of symmetries \cite{pagels-stokar, ball-chiu}.
The subtle effect is related with the fact that the fermion propagator $S^{-1}(p)=\slashed p - \Sigma(p)$ with the momentum-dependent mass
has to be handled in the WT identities very carefully. The new terms which they induce are consequently characterized by the 'derivatives' of $\Sigma_f(p^2)$.\\

\subsection{1. The Ward-Takahashi identities}

The electromagnetic WT identities for the proper vertices depicted in Figs.1,2 read as \cite{pagels-stokar}, \cite{delbourgo}, \cite{benes-hosek}
\begin{widetext}
\begin{eqnarray*}
q_{\mu}\Gamma^{\mu}(p',p) &=& S^{-1}(p') - S^{-1}(p)\\
k_{1\mu}\Gamma^{\mu \nu}(k_1, k_2, p', p) &=& -\Gamma^{\nu}(p', p'+k_2) + \Gamma^{\nu}(p-k_2, p)\\
k_{2\nu}\Gamma^{\mu \nu}(k_1, k_2, p', p) &=& -\Gamma^{\mu}(p', p'+k_1) + \Gamma^{\mu}(p-k_1, p)
\end{eqnarray*}
\end{widetext}
The proper vertices themselves satisfy the WT identities, have no unwanted kinematic singularities and together with  $S^{-1}(p)=\slashed p - \Sigma(p)$
define a new dynamical perturbation theory (DPT) \cite{pagels-stokar}. They have the form
\begin{widetext}
\begin{eqnarray*}
\Gamma^{\mu}(p',p) &=& [\gamma^{\mu} - (p' + p)^{\mu}\Sigma'(p',p)]\\
\Gamma^{\mu \nu}(k_1,k_2,p',p) &=& [-2g^{\mu \nu}\Sigma'(p',p) +\\ &&(k_1-2p)^{\mu}(k_2+2p')^{\nu}\Sigma''(p-k_1,p',p) + (k_1+2p')^{\mu}(k_2-2p)^{\nu}\Sigma''(p-k_2,p',p)]
\end{eqnarray*}
\end{widetext}
The 'derivatives' are defined as
%\begin{widetext}
\begin{eqnarray*}
\Sigma'(p_1,p_2) &\equiv& \frac{\Sigma(p_1) - \Sigma(p_2)}{p_1^2 - p_2^2}\\
\Sigma''(p_1, p_2, p_3) &\equiv& \frac{\Sigma'(p_1, p_3) - \Sigma'(p_2, p_3)}{p_1^2 - p_2^2}
\end{eqnarray*}
%\end{widetext}
\begin{figure}[t]
\begin{center}
\includegraphics[width=1\linewidth]{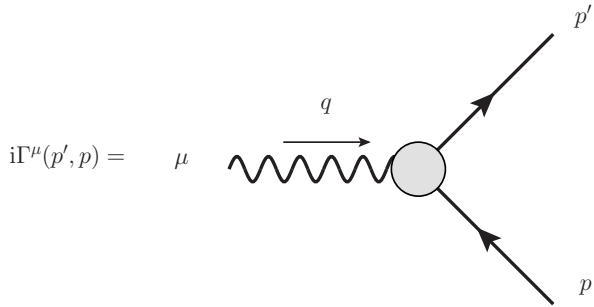}
\end{center}
\caption{Proper vertex $\Gamma^\mu(p^\prime,p)$, with $q=p^\prime-p$.}
\label{fig_vertex3}
\end{figure}

\begin{figure}[t]
\begin{center}
\includegraphics[width=1\linewidth]{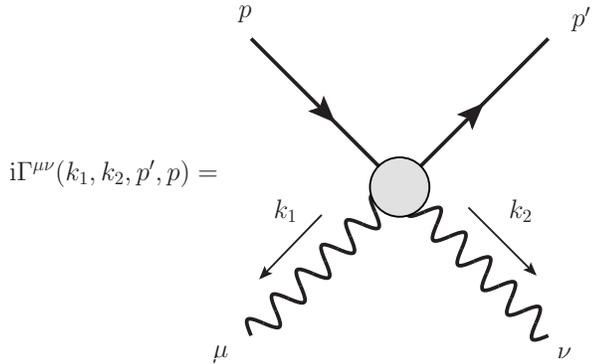}
\end{center}
\caption{Proper vertex $\Gamma^{\mu\nu}(k_1,k_2,p^\prime,p)$ with $p-k_1=p^\prime+k_2$.}
\label{fig_vertex4}
\end{figure}

It is important to emphasize that in the vertex $\Gamma^{\mu}$ the bare part is chirality-conserving whereas the induced one is chirality-changing. The vertex
$\Gamma^{\mu \nu}$ is entirely induced and hence it is chirality-changing. Because these generically new vertices emerge only in the electromagnetic
interactions our conjecture refers only to them. For an illustration we consider explicitly only the $\Gamma^{\mu}$.\\

\subsection{2. Illustrative pole-mass splitting}

In the Landau gauge and in the lowest order DPT the massless photon contribution $\delta_i\Sigma(p^2)$ to the fermion propagator
$S(p)=(\slashed p - \Sigma(p^2))^{-1}$ with the vertex $\Gamma^{\mu}$ is \cite{difference}
\begin{widetext}
\begin{eqnarray*}
\delta_i\Sigma(p^2) = e^2 Q_i^2 \int \frac{d^4k}{(2\pi)^4}\frac{1}{(p-k)^2 [k^2 + \Sigma^2(k^2)]}\{3\Sigma(k^2) + 4\frac{\Sigma(p^2) - \Sigma(k^2)}
{p^2 - k^2}\frac{p^2k^2 - (p.k)^2}{(p - k)^2}\}
\label{Sigmai}
\end{eqnarray*}
\end{widetext}
The formula is valid for every family $f$, and the index $f$ is therefore omitted for simplicity. The index $i$ runs over the charged lepton with $Q=-1$, the quark
with $Q=2/3$, and the quark with $Q=-1/3$ in a chosen family.
We define $\delta_i \Sigma(p^2) \equiv \delta_i \Sigma^a(p^2) + \delta_i \Sigma^b(p^2)$. The first term $(a)$ in curly brackets is the contribution from the bare
chirality-conserving fermion-photon vertex,  the term $(b)$ is the contribution of the new, induced chirality-changing vertex.

Using the explicit form $\Sigma (p^2)=m^2/p$ we get
\begin{widetext}
\begin{equation}
\delta_i \Sigma^a(p^2) = \frac{3\alpha Q_i^2}{2\pi}\frac{m^2}{p} +  \frac{3\alpha Q_i^2}{2\pi}m \frac{1}{2\surd 2}(F -  G + \frac{\pi}{2}) -
\frac{3\alpha Q_i^2}{2\pi}\frac{m^3}{p^2} \frac{1}{2\surd 2}(F + G)
\end{equation}
\end{widetext}
\begin{widetext}
\begin{equation}
\delta_i \Sigma^b(p^2) = \frac{3\alpha Q_i^2}{2\pi}\frac{m^2}{p}\frac{1}{2}(1 - H) - \frac{3\alpha Q_i^2}{2\pi}m \frac{1}{2\surd 2}(F -  G) -
\frac{3\alpha Q_i^2}{2\pi} \frac{p^2}{m} \frac{1}{2\surd 2}(F + G) - \frac{3\alpha Q_i^2}{2\pi} \frac{m^4}{p^3} \frac{1}{2\surd 2}[\frac{\pi}{2} - {\rm arctan} (p/m)^2]
\end{equation}
\end{widetext}
where $F,G,H$  are dimensionless slowly varying functions
\begin{equation*}
F = \tfrac{1}{2}{\rm ln} \tfrac{(p/m + \tfrac{1}{\surd 2})^2 + \tfrac{1}{2}}{(p/m - \tfrac{1}{\surd 2})^2 + \tfrac{1}{2}},\phantom{bbb}  G = {\rm arctan} \tfrac{(p/m)\surd 2}{1 - (p/m)^2}
\end{equation*}
\begin{equation*}
H = \tfrac{1}{2}\{\tfrac{1}{4}{\rm ln}(1 + (\tfrac{p}{m})^4) - \tfrac{(p/m)^4}{1 + (p/m)^4} {\rm ln} (\tfrac{p}{m})\}
\end{equation*}
which should not influence too much the leading behavior of $\delta_i \Sigma^a(p^2)$ and $\delta_i \Sigma^b(p^2)$.\\

In general, the masses $m_i$ of the electrically charged fermions are given by the poles of the propagators $S_i(p)=(\slashed p - \Sigma_i(p^2))^{-1}$
with $\Sigma_i(p^2) \equiv \frac{m^2}{p} + \delta_i \Sigma(p^2)$ i.e., by solving the algebraic equation
\begin{equation}
m_i = \frac{m^2}{m_i} + \delta_i\Sigma(p^2=m_i^2).
\end{equation}\\

(1) The contribution
\begin{equation*}
\frac{3\alpha Q_i^2}{2\pi}\frac{m^2}{p}
\end{equation*}
to $\Sigma(p^2)=m^2/p$ is by definition of order $\alpha$ i.e.,
\begin{equation*}
(\tfrac{m_i}{m})^2 = 1 + \tfrac{3\alpha Q_i^2}{2\pi}
\end{equation*}\\

(2) The constant contribution
\begin{equation*}
\frac{3\alpha Q_i^2}{2\pi}m
\end{equation*}
to $\Sigma(p^2)=m^2/p$ is also of order $\alpha$,
\begin{equation*}
(\tfrac{m}{m_i})^2 = 1 - \tfrac{3\alpha Q_i^2}{2\pi}
\end{equation*}\\

(3) The contribution
\begin{equation*}
-\frac{3\alpha Q_i^2}{2\pi}\frac{m^3}{p^2}
\end{equation*}
to $\Sigma(p^2)=m^2/p$ is determined by solutions of the cubic equation $\tfrac{3\alpha Q_i^2}{2\pi}x^3 - x^2 + 1 = 0$ for $x=m/m_i$.
There is no acceptable solution since all three real roots are negative.\\

(4) The contribution
\begin{equation*}
-\frac{3\alpha Q_i^2}{2\pi}\frac{p^2}{m}
\end{equation*}
to $\Sigma(p^2)=m^2/p$ is determined by solutions of the cubic equation $x^3 - x - \tfrac{3\alpha Q_i^2}{2\pi} = 0$ for $x=m/m_i$.
The positive root is $\tfrac{m}{m_i} = 1 + O(\alpha)$; the other real roots are negative.\\

(5) The contribution
\begin{equation*}
-\frac{3\alpha Q_i^2}{2\pi}\frac{m^4}{p^3}
\end{equation*}
to $\Sigma(p^2)=m^2/p$ is determined by solutions of the bi-quadratic equation $\tfrac{3\alpha Q_i^2}{2\pi}x^4 - x^2 + 1 = 0$ for $x^2=(m/m_i)^2$.
One solution is of the form $(m/m_i)^2 = 1 + O(\alpha)$, the other is inversely proportional to $\alpha$,
\begin{equation*}
(m/m_i)^2 = \frac{2\pi}{3\alpha Q_i^2}
\end{equation*}
This explicitly demonstrates that the fermion pole-mass splitting due to new $\Sigma(p^2)$-induced chirality changing
fermion-photon vertex $\Gamma^{\mu}$ can be non-analytic in the fermion electric charges i.e., large.

The numerical analysis of the full $\Sigma_i(p^2)$ confirms the simple analysis presented above: The standard QED vertex $\gamma^{\mu}$ generates only
the perturbative mass splitting $m/m_i \sim 1 + O(\alpha Q_i^2)$. With the complete vertex $\Gamma^{\mu}$ there are two solutions:
(i) The perturbative mass splitting. (ii) The large but nonrealistic mass splitting  in which the behavior of point 5,
$(m/m_i)^2 = \frac{2\pi}{3\alpha Q_i^2}$ dominates: $m_i^2/m_j^2=Q_i^2/Q_j^2$.

We believe that the illustrative computation presented above supports our conjecture: With the 'right' functional form of $\Sigma(p^2)$
and with both $\Gamma^{\mu}$ and $\Gamma^{\mu \nu}$ fermion-photon vertices {\it the observed large UV finite fermion mass splitting within families is due to different electric charges}.

\section{IV. Conclusion}
Provided the property $M_{fR} \gg m_f$ is generic the replacement of the Higgs sector of the Standard model by QFD results in
natural, predictive and rigid scenario:\\

First, the model is the natural quantum field theory of seesaw: Introduction of one triplet of sterile
right-handed neutrinos for anomaly freedom is the necessary quantum-field-theoretic requirement. Prediction
of the neutrino mass spectrum computed solely by QFD is in accord with the effective field theory prediction of Steven Weinberg \cite{weinberg-majorana}:
{\it Three active neutrinos are the massive, extremely light {\it Majorana} fermions.}\\

Second, {\it there is no genuine Fermi scale} in the model \cite{pagels-stokar}. Masses of $W$ and $Z$ bosons as well as the masses of the composite Higgs bosons
$h, h_3, h_8$ are the consequence of the QFD-generated Dirac masses $m_f$
exponentially light with respect to the huge QFD scale $\Lambda$, the only scale in the game. Assuming for
simplicity that the sum rule for the $W,Z$ masses is saturated by the heaviest Dirac mass $m_3=\Lambda \phantom{b} \rm exp \phantom{b}
(-4\pi^2/g_{88})$ we get  from the formula \cite{hosek}
\begin{equation*}
m_W^2=\tfrac{1}{4}g^2\tfrac{5}{4\pi} m_{3}^2
\end{equation*}
the estimate $m_3 \sim 390 \phantom{b}\rm GeV$. Because of the subsequent QED pole mass splitting $m_3$ is the unobservable Dirac neutrino mass entering seesaw.
The heaviest active Majorana neutrino mass taken as $m_{\nu_3}=m_3^2/\Lambda \sim  \rm \phantom{b} 1.6\phantom{b} eV$  implies $\Lambda \sim 10^{14} \phantom{b} \rm GeV$.\\

Third, {\it the masses $m_f$ and $M_{fR}$ computed by QFD uniquely fix the pattern of spontaneously broken symmetries}.
By Goldstone theorem there is the fixed pattern of masses of the gauge bosons, of the composite pseudo NG bosons, and
also the fixed pattern of the composite Higgs particles. In particular, the robust prediction is {\it the existence of two new additional
composite Higgs particles $h_3$ and $h_8$ with masses at the Fermi scale}. Because of their peculiar flavor-dependent
couplings with the SM fermions \cite{hosek} they should not escape detection. It is perhaps noteworthy
that their existence is directly related with three fermion families: Regular representation of $SU(3)$ has two
diagonal generators.\\

Fourth, computation of the {\it ultimately parameter-free fermion mass spectrum} proceeds in two stages.
(1) The strongly coupled QFD generates the calculable $M_{fR}$ and $m_f$, and is responsible for the formation of the 'would-be' NG bosons,
of the pseudo-NG bosons and of the composite Higgs-like particles. (2) The weakly coupled QED
with new $\Sigma(p^2)$-induced fermion-photon vertices is responsible for the charged-lepton and quark mass splitting in terms of the fermion electric charges.
It is important that the QED fermion mass splitting does not imply any additional
spontaneous symmetry breaking. We expect that, as in the SM \cite{carter-pagels}, the main properties of the gauge boson masses, of the pseudo NG bosons and of the
composite Higgs-like particles will be only perturbatively modified by the gauge electroweak corrections.\\

Fifth, QFD offers a {\it natural candidate for dark matter} as a straightforward analog of luminous matter, the colorless
QCD-composite nucleons $N \sim \epsilon_{abc} q^a q^b q^c$ forming the stable nuclei.
A suggestion is that the mass of the dark Universe is made of heavy sterile neutrinos in the form of the QFD-composite fermions
$S \sim \eta_{abc} \nu_R^a \nu_R^b \nu_R^c$. Here $\eta$ is the favorite Clebsch-Gordan coefficient of $SU(3)$ discussed below:
The QCD is confining in the infrared i.e, the nucleon is the color singlet. This is heuristically supported by the properties of gluon exchanges:
In the decomposition $3 \times 3 \times 3 = (3^{*} + 6) \times 3 = (3^{*} \times 3) + (6 \times 3) = 1 + 8 +10 + 8$ the diquark is formed
by an attraction in $3^{*}$ channel, and bound with $3$ into the color singlet. The QFD is also strongly coupled in the infrared
but, by assumption, self-broken in a definite pattern: Two flavor triplets of sterile neutrinos attract each other in flavor sextet (in this channel
there are the composite 'would-be' NG bosons). The di-neutrino sextet then forms with the third neutrino either the flavor octet or the flavor decuplet.
In the resulting composite 'dark' fermion $S$ the spins and angular momenta of its sterile neutrinos are fixed in accord with the Pauli principle.\\

The scenario is rigid. First, its free parameters are the two electroweak gauge coupling constants $g, g'$
and the QFD scale $\Lambda$. (QCD stays dormant in what we are doing.) Second, its non-perturbative properties are the consequences
of spontaneously generated $\Sigma_f(p^2)$.

An optimistic side of the model is that its global properties
like the masses of gauge bosons or the properties of the Higgs bosons are given by the integral formulas like the Pagels-Stokar
formula for the gauge boson masses or the triangle loop formulas for the Higgs decay rates \cite{benes-hosek}, in which the detailed behavior of $\Sigma(p^2)$ smears out.

A pessimistic side of the model is that for explaining the observed pattern of the fermion mass spectrum
the detailed functional form of  $\Sigma(p^2)$ is the necessity. Because it is the result of formidable
strong-coupling computation we expect that in this most interesting respect the model is sentenced to stay only semi-quantitative.\\

\begin{acknowledgements}
I am indebted to my colleagues Adam Smetana and Petr Bene\v s for many interesting discussions and for generous help with the manuscript.
The work on this project has been supported by the grant LG 15052 of the Ministry of Education of the Czech Republic.
\end{acknowledgements}


\begin{thebibliography}
\bibitem{argument} The fermion mass terms themselves are super-renormalizable
but, being non-invariant they break down the gauge symmetry in loops. This implies the necessity of introducing the
non-invariant gauge boson mass counter-terms i.e., the unitarity-violating hard gauge boson masses.
\bibitem{difference} We differ from \cite{pagels-stokar} by the factor of 2
in the second term in curly brackets.
\end{thebibliography}
\end{document}